\documentclass[a4paper,11pt]{article}
\usepackage{jinstpub} 
\usepackage{hyperref}
\usepackage{listings}
\usepackage{xcolor}
\usepackage{multirow}

\definecolor{keywords}{RGB}{0,125,0}
\definecolor{comments}{RGB}{0,100,113}
\definecolor{red}{RGB}{160,0,0}
\definecolor{green}{RGB}{0,150,0}

\title{\boldmath A simulation framework for SiPMs}

\author[a]{J. Peña-Rodríguez}
\author[a]{J. Förtsch}
\author[a]{C. Pauly}
\author[a]{K.-H. Kampert}

\affiliation[a]{Department of Physics, University of Wuppertal, Gaußstraße 20, Wuppertal, D-42119, Germany}

\emailAdd{penarodriguez@uni-wuppertal.de}

\abstract{We present a Python module for simulating Silicon Photo-Multipliers. This module allows users to perform noise analyses: Dark Count Rate, crosstalk, and afterpulsing. Furthermore, the simulation framework novelty is the capability of simulating assemblies of SiPM arrays for large area detectors like Ring Imaging Cherenkov detectors, Cherenkov Telescopes, Positron Emission Tomography, and any detector using them. Users can simulate ring- or shower-like-shaped signals based on the expected number of photons generated by the source. We validate the performance of the simulation module with data from four different SiPM: Broadcom AFBR-S4N66P024M, Hamamatsu S14160-636050HS, Onsemi MICROFC-60035, and FBK NUV-HD3.}


\arxivnumber{2411.16710v1} 

\begin{document}
\maketitle
\flushbottom

\section{Introduction}
\label{intro}

For decades, high-energy physics and astrophysics experiments have been governed by using photomultiplier tubes (PMTs) as photon detection systems. Low thermal noise, a high photoelectron multiplication factor, and radiation hardness are the main strengths that make PMTs well suited for most of these experiments. However, experiments are moving forward, searching for new photodetection technologies to explore more accurate timing, spatial, and amplitude resolutions. Silicon photomultipliers (SiPM) measure light intensities down to single-photon level with picosecond timing precision, photodetection efficiencies reaching up to 60\%, and immunity against magnetic fields.

Drawbacks of SiPMs, in particular in comparison to Multi-anode Photomultipliers (MAPMTs), include high Dark Count Rates (DCR), temperature dependency, radiation susceptibility, and in case of SiPM arrays slightly lower fill-factor. Typical DCR of SiPM at room temperature are five orders of magnitude larger, $\sim10^5$\,Hz/mm$^2$ in comparison to $\sim$\,Hz/mm$^2$ in case of modern MAPMTs.

Implementation of SiPM-based systems has been tested in Imaging Air Cherenkov Telescope (IACT) for high energy gamma ray detection \cite{Anderhub2011, Neise2017, Depaoli2023, Schwanke2015}, Ring Imaging Cherenkov (RICH) detectors \cite{Basso2024, Preghenella2023}, and Positron Emission Tomography (PET) \cite{Lecoq2021}. They are usually applied in triggered readout systems, using e.g.\ majority voting and pulse leading edge coincidences.

The feasibility of using SiPM in large area detectors needs to be validated before starting major hardware development, and Monte Carlo simulations of the expected detector response represent a useful tool for evaluation, development, and further tuning. There exist software for SiPM signal/noise simulation based on electrical or Monte Carlo models using SPICE, GATE or GEANT4 \cite{Mehadji2022,Acerbi2019, Rosado2015,Niggemann2015,Jha2013}. However, these tools are addressed to simulate individual SiPM but not large-area detection arrays.

In this work, we present a Python framework for modeling detector signals and for analyzing detection systems based on SiPMs. The framework simulates uncorrelated (DCR) and correlated (crosstalk and afterpulse) noise sources, giving realistic data for testing triggering approaches, signal recognition algorithms, photosensor evaluation, and detector response. The user can inject typical RICH/IACT signals (rings or traces) to the simulated detector for signal-to-noise ratio (SNR) analysis.

We validated the simulation tool by comparing with data of four commercially available SiPM (Broadcom AFBR-S4N66P024M, Hamamatsu S14160-6050HS, Onsemi MICROFC-60035, and FBK NUV-HD3) under different operation conditions of bias voltage and temperature. Section \ref{sec::sipm} shows the SiPM pulse shape model and section \ref{sec::noise} covers the noise modelling. In section \ref{sec::validation}, we present the SiPM model validation. Section \ref{sec::rich} illustrates application of the framework for simulation of two cases: a RICH detector and an IACT.

\section{SiPM signal modeling}
\label{sec::sipm}

\begin{figure}[ht]
    \centering
    \includegraphics[width=8cm]{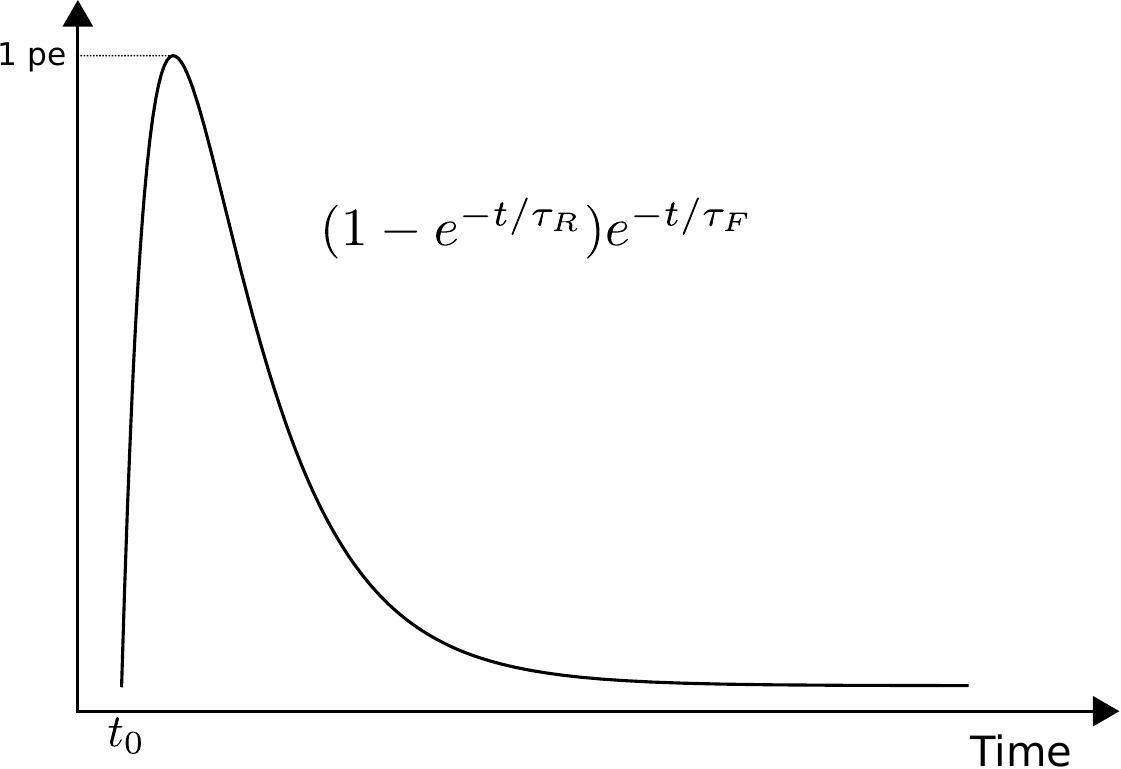}
    \caption{SiPM pulse model. The signal rise time depends on the microcell discharge resistance $R_S$ and the junction capacitance $C_J$, while the fall time depends on the quenching resistance $R_Q$ and the junction capacitance $C_J$. The Geiger discharge starts at $t_0$.}
    \label{fig:pulse}
\end{figure}

The SiPM pulse shape is a conjunction of multiple successive processes starting with the incident photon absorption in the silicon crystal, the electron avalanche development inside the substrate, the avalanche quenching, and the diode capacitance recharging. The signal can simply be characterized by a leading and falling edge region (recovery) \cite{Acerbi2019,Gundacker2020,Mehadji2022}, described by 

\begin{equation}
    A(1-e^{-t/\tau_R})e^{-t/\tau_F}
\end{equation}

\noindent and shown in Fig.\,\ref{fig:pulse}. Here, $\tau_R = R_SC_J$ is the rise time constant with the microcell resistance $R_s$ and junction capacity $C_J$. $\tau_F = R_QC_J$ is the fall time constant with the quenching resistance $R_Q$. $A$ is the amplitude parameter. The signal amplitude is modeled by a Gaussian distribution with mean 1\,pe and a default standard deviation of 0.1\,pe (which can be adjusted).

\section{Correlated and uncorrelated noise modelling}
\label{sec::noise}

\subsection{Dark Count Rate}
SiPM noise is mainly generated by charge carriers (electron/hole) released due to thermal excitation inside the silicon substrate. These charge carriers create spurious current pulses, dark counts, even in the absence of light, that are indistinguishable from the photon-generated ones. The Dark Count Rate (DCR) follows a Poisson distribution with the exponential constant equal to the DCR value \cite{Hamamatsu2024,Montaruli2021,Acerbi2019}.

The distribution function of dark counts is defined by

\begin{equation}
    f_{DCR}(t) = \frac{1}{\tau} \: e^{-t/\tau},
\end{equation}

\noindent where $t$ is the time difference between two consecutive dark counts and $\tau=1/\text{DCR}$ with DCR measured in Hz.\\

\subsection{Afterpulsing}

Correlated SiPM noise refers to noise sources closely related with a primary event (dark count or photon-event), and mainly originates from two effects: afterpulsing and crosstalk.
Afterpulsing is due to charge carriers trapped in silicon defects during the avalanche development and being released later, generating a new avalanche or due to photons, created during the primary avalanche,  convert in the non-depleted region into a minority charge carrier that diffuses into the high-field region of the same microcell and produces a Geiger discharge there. 

The afterpulse charge increases with the relative time between the primary event and the instant when the trapped charge carrier is released, because of the time needed to recharge the pn junction capacity and restore the bias (over)voltage. The afterpulse amplitude is defined as
\begin{equation}
    A_{AP}(t) = 1 - e^{-t/\tau_{rec}},
\end{equation}

where $t$ is the time difference between the main pulse and the afterpulse and $\tau_{rec} = \tau_F = R_Q C_J$ is the SiPM recovery time \cite{Rosado2015,Montaruli2021}. The distribution of afterpulses is modeled as

\begin{equation}
    f_{AP}(t) = \frac{\tau_{rec} + \tau_{AP} }{\tau_{AP}^2} \left( 1 - e^{-t/\tau_{rec}} \right) \: e^{-t/\tau_{AP}},
\end{equation}

where $\tau_{AP}$ is the mean releasing time of trapped charge carriers and $\tau_R < \tau_{AP} < \tau_{rec}$ \cite{Nagy2014}.

\subsection{Crosstalk}

The optical crosstalk process involves photons that are emitted during the primary avalanche (dark count or photon) and get absorbed in neighboring cells surrounding the primary cell. 
This re-absorption creates a new avalanche and, as a consequence, a new current pulse. The crosstalk amplitude directly depends on the number of activated neighboring cells. The crosstalk probability is modeled as a geometric chain process \cite{Razeto2024,Vinogradov2012,Gallego2013}, where a Poisson distribution models the primary avalanche source (mainly DCR) and the chain process is:

\begin{equation}
    P_{Npe} = (\alpha P_{CT})^N,
\end{equation}
where $P_{Npe}$ is the probability of crosstalk with amplitude above $N$\,pe after a primary avalanche, $P_{CT}$ the total crosstalk probability, and $\alpha$ is a scaling factor obtained from model fitting.

\begin{figure}[ht]
    \centering
    \includegraphics[width=9cm]{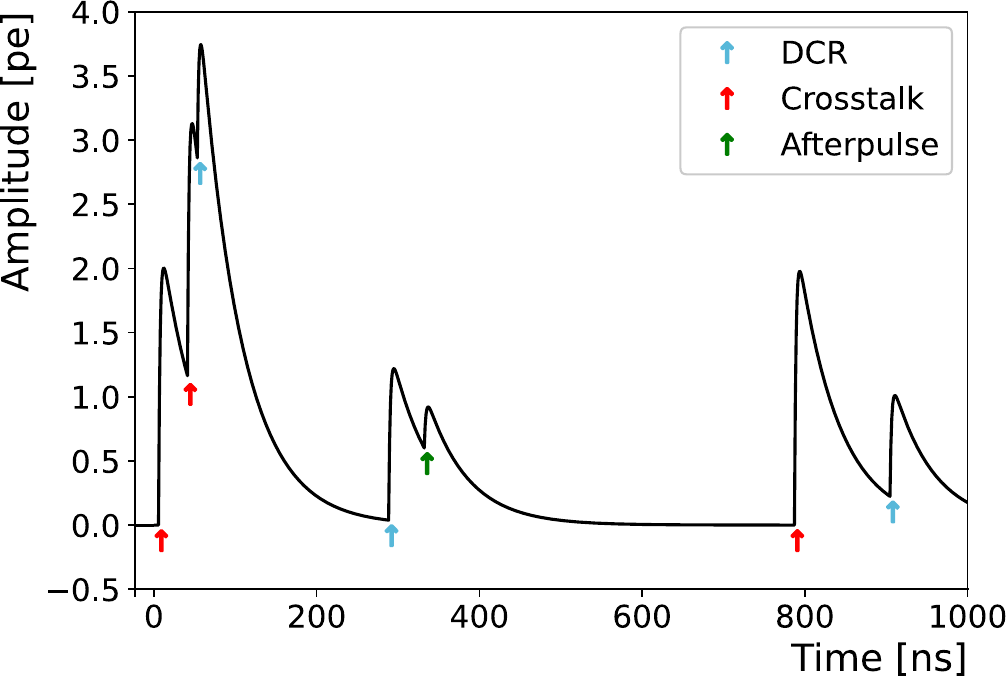}
    \caption{Example of a simulated SiPM signal. Components: DCR (purple arrow), crosstalk (red arrow), afterpulse (green arrow). The time bin is 100\,ps.}
    \label{fig:signal}
\end{figure}

The SiPM signal waveform is generated by summing all signal components (DCR, crosstalk or afterpulse) in a given time bin as is shown in Fig.\,\ref{fig:signal}. The time bin is set by the user depending on the time resolution.

\section{Signal analysis}
\label{sec::analysis}
The noise quantification of the real and simulated data was performed following the same procedure. However, the real SiPM signals were first normalized to photoelectrons for direct comparison with the simulated data. The photoelectron value was determined by obtaining the amplitude spectrum of the SiPM signals, fitting a Gaussian distribution to each spectrum peak, determining the numerical difference between the mean values of the consecutive distributions, and averaging all the differences.

The signal analysis extracted the time difference between consecutive peaks and the peak amplitude value. A peak finder algorithm detects the signal peaks, where a peak is defined as any sample (or flat region) whose two direct neighbors have a smaller amplitude. We fixed the simulation time window equal to the data time window in order to directly compare the model and data results.

Additionally, the DCR was calculated by,

\begin{equation}
    DCR = \frac{N_{th}}{T},
\end{equation}
where $N_{th}$ is the number of peaks with amplitude above a set threshold and $T$ is the temporal size of the analyzed signal.

\section{Model validation}
\label{sec::validation}

We validated our model by comparing it with data from four SiPM: Broadcom AFBR-S4N66P024M, Hamamatsu S14160-6050HS, Onsemi MICROFC-60035, and FBK NUV-HD3. The first three SiPMs were characterized in our laboratory while the characterization of the FBK NUV-HD3 SiPM was made by the pSCT experiment \cite{Ambrosi2023}.

The SiPM (AFBR-S4N66P024M, S14160-6050HS, or MICROFC-60035) was placed in a light-tight dark box where the temperature (measured with an uncertainty of 0.1$^{\circ}$C) and bias voltage were controlled. A picosecond laser PLDD-20M stimulated the SiPM at a rate of 100\,kHz. The SiPM readout consisted of a Class-A RF amplifier based on the BFU760F transistor and a pole-zero cancellation circuit for pulse shortening (eliminating pile-up effects). The data was recorded by an oscilloscope at 1\,GHz sampling frequency. The SiPM parameters are shown as a Table in \ref{tab:sipm}.

\begin{table}[h!]
    \caption[xx]{SiPM parameters. FBK NUV-HD3 parameters were obtained at 20$^{\circ}$/5.5V$_{ov}$\cite{Ambrosi2023}.}\label{tab:sipm}
    \begin{center}
    \begin{tabular}{|l|c|c|c|c|} \hline 
        \textbf{SiPM} & \textbf{MICROFC-60035} & \textbf{S14160-6050CS} & \textbf{AFBRS4N66P024} & \textbf{FBK NUV-HD3} \\  \hline 
        \textbf{Vendor} & OnSemi & Hamamatsu & Broadcom & FBZ \\ 
        \textbf{Active Area (mm$^2$)} & $6 \times 6$  & $6 \times 6$  & $6 \times 6$ & $6 \times 6$ \\
        \textbf{Pixel pitch ($\mu$m)} & 35 & 50  & 40 &  40 \\
        \textbf{Number of cells} & 18980 & 14331 & 22428  & $\sim$22500 \\
        \textbf{Fill factor($\%$)} & 64 &  74 & $\sim$80 &  $\sim$85 \\
        \textbf{$V_{br}$ (V)} & 24.5 & 38 & 45 & 26.5 \\
        \textbf{PDE ($\%$)} & 41 & 50  & 63 & 55.7 \\
        \textbf{Wav. peak (nm)} & 420  & 450  & 420 & 400  \\
        \textbf{Gain ($\times10^6$)} & 3 & 2.5 & 7.3 & 4  \\
        \textbf{DCR (kHz/mm$^2$)} & 95 & 100 & 125 & 72 \\
        \textbf{Crosstalk ($\%$)} & 7 & 7 & 23 & 25  \\
        \textbf{Afterpulsing ($\%$)} & 0.2 & -  & 1 & - \\ \hline
    \end{tabular}
    \end{center}
\end{table}

Based on the SiPM pulse-shape analysis, we measured fall times of 55\,ns for AFBR-S4N66P024M, $\sim$60\,ns for S14160-6050HS, and 95\,ns for MICROFC-60035. In our set-up the pulse shape is modified by the shortening circuit obtaining a rise time of 2\,ns and a fall time of 3\,ns. Noise curves (DCR, crosstalk, and afterpulsing) depending on temperature and bias voltage for all the SiPM were characterized taking into account methodologies reported in \cite{Ghassemi2018}. The model performance was evaluated by using four parameters: the pulse peak-amplitude spectrum, DCR vs.\ threshold, amplitude vs.\ relative time between consecutive pulses (inter-time), and relative time distribution.

\begin{figure}[ht]
    \centering
    \includegraphics[width=7cm]{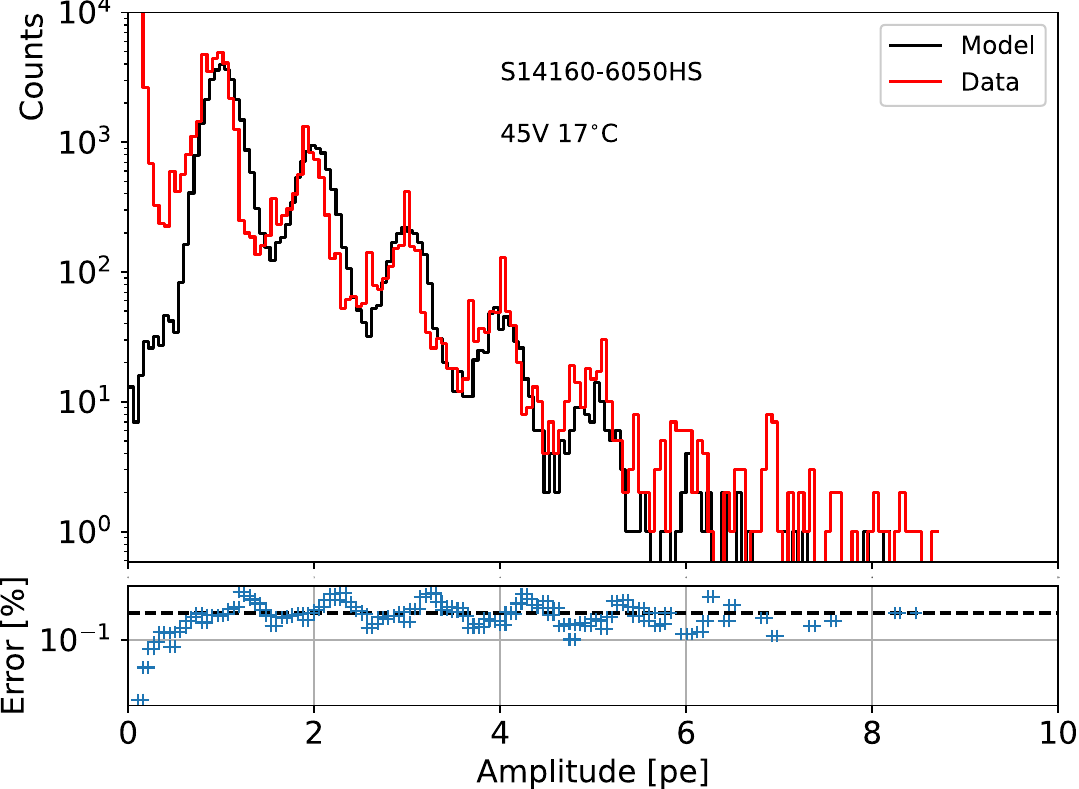}
    \includegraphics[width=7cm]{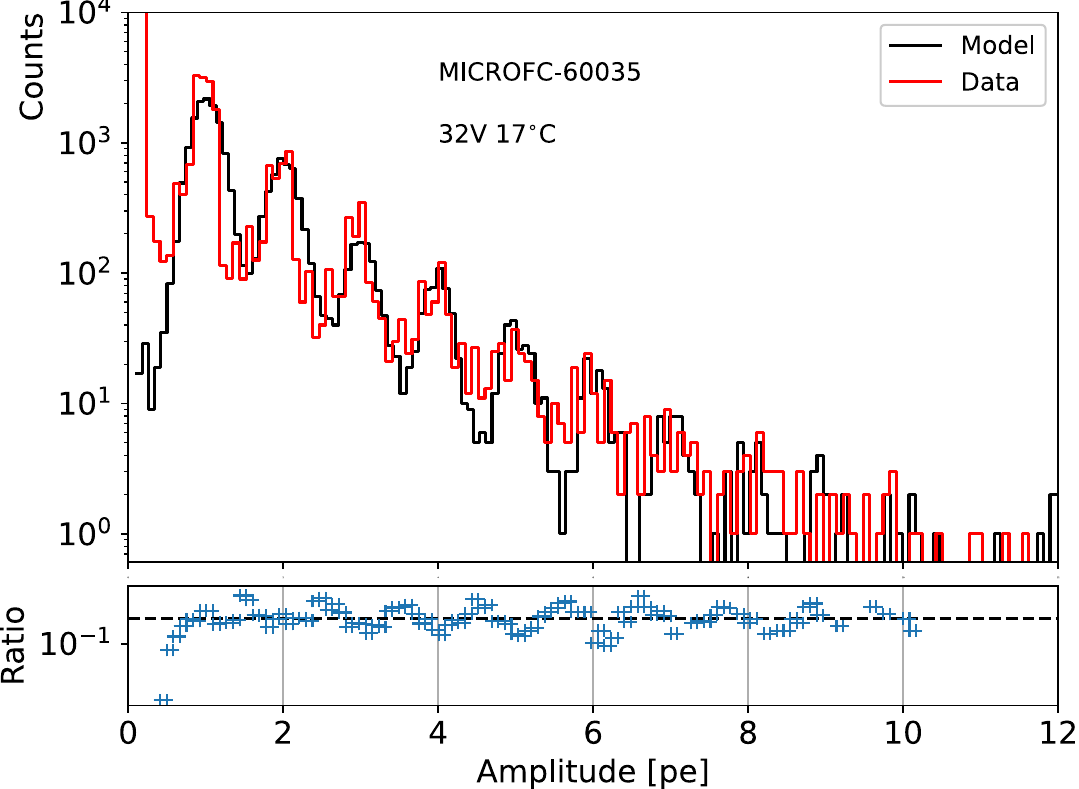}
    \caption{Amplitude spectrum comparison between the model (black-line) and data (red-line) for the S14160-6050HS at 45V/17$^{\circ}$C (left) and MICROFC-60035 at 32V/17$^{\circ}$C (right). The first data peak is the pedestal.}
    \label{fig:peak_spectrum}
\end{figure}

Figure \ref{fig:peak_spectrum} shows the peak spectra of the S14160-6050HS and MICROFC-60035 operating in dark conditions. The S14160-6050HS was set to 45V/17$^{\circ}$C where the DCR is 340.5\,kHz/mm$^2$ and the crosstalk is 22$\%$. The MICROFC-60035 was set to 32V/17$^{\circ}$C with a DCR of 166.3\,kHz/mm$^2$ and a crosstalk of 31$\%$. The data spectrum (red line) was normalized to 1\,pe, and the first maximum close to 0 is the pedestal caused by the baseline noise. The model describes the data behavior around the photoelectron peaks, where the model/data ratio is around one. However, in the spectrum valleys, the ratio increases, showing some differences between the model and the data. The signal noise slightly enlarges the amplitude, filling the spectrum gaps between photoelectron levels. The simulation framework does not model the signal noise.

\begin{figure}[ht]
    \centering
    \includegraphics[width=6.5cm]{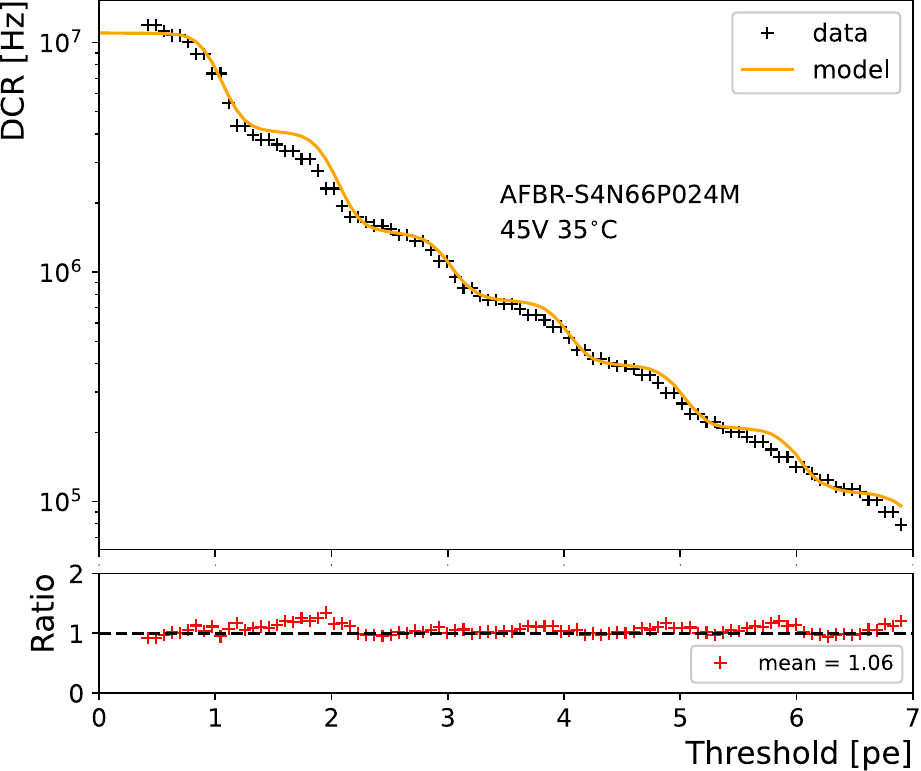}
    \includegraphics[width=6.5cm]{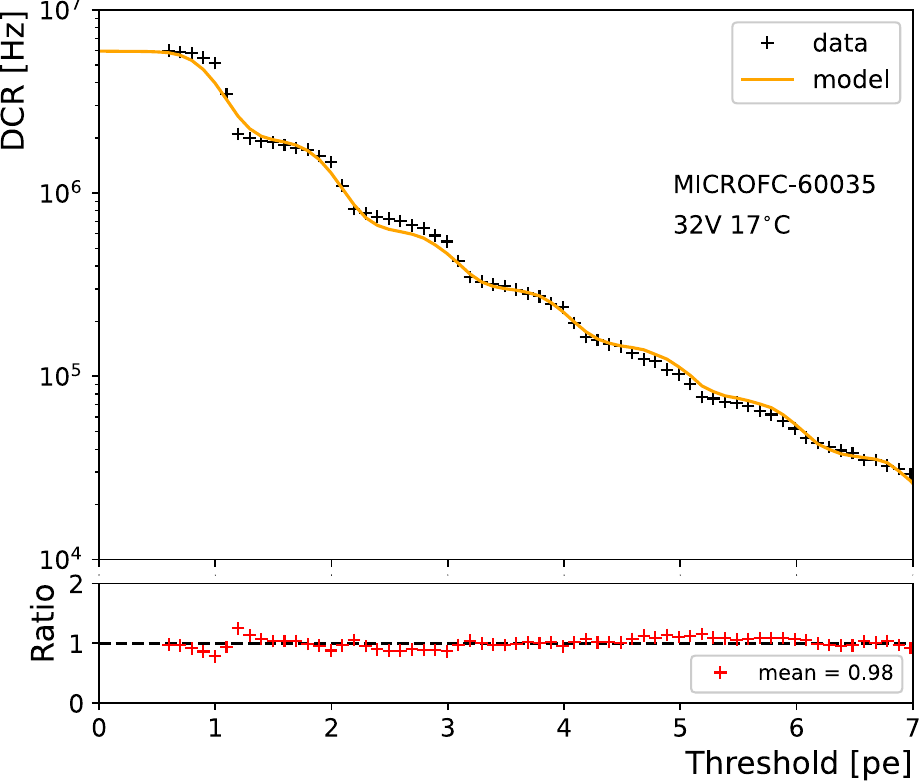}
    \caption{DCR vs.\ threshold curve comparison between the model (orange-line) and data (black-cross)  for the AFBR-S4N66P024M at 45V/35$^{\circ}$C (left) and MICROFC-60035 at 32V/17$^{\circ}$C (right).}
    \label{fig:dcr_th}
\end{figure}

The DCR vs.\ threshold curve represents the estimated noise signal rate as function of detection threshold. This parameterization helps to configure triggering systems depending on the amplitude (in terms of number of photo electrons) the expected signal has. The DCR vs.\ threshold curve for the AFBR-S4N66P024M (45V/35$^{\circ}$C) and the MICROFC-60035 (32V/17$^{\circ}$C) are shown in Fig.\,\ref{fig:dcr_th}. We calculated the model/data ratio for evaluating the model performance. The AFBR-S4N66P024M DCR was set to 11.23\,MHz (0.5\,pe level) with a crosstalk probability of 34\%. The model prediction follows the data behavior resulting in a model/data ratio of 1.06. In the second case, the MICROFC-60035 DCR was set to 5.98\,MHz with a crosstalk of 31\%. The average of the ratio model/data was 0.98.

\begin{figure}[ht]
    \centering
    \includegraphics[width=6.5cm]{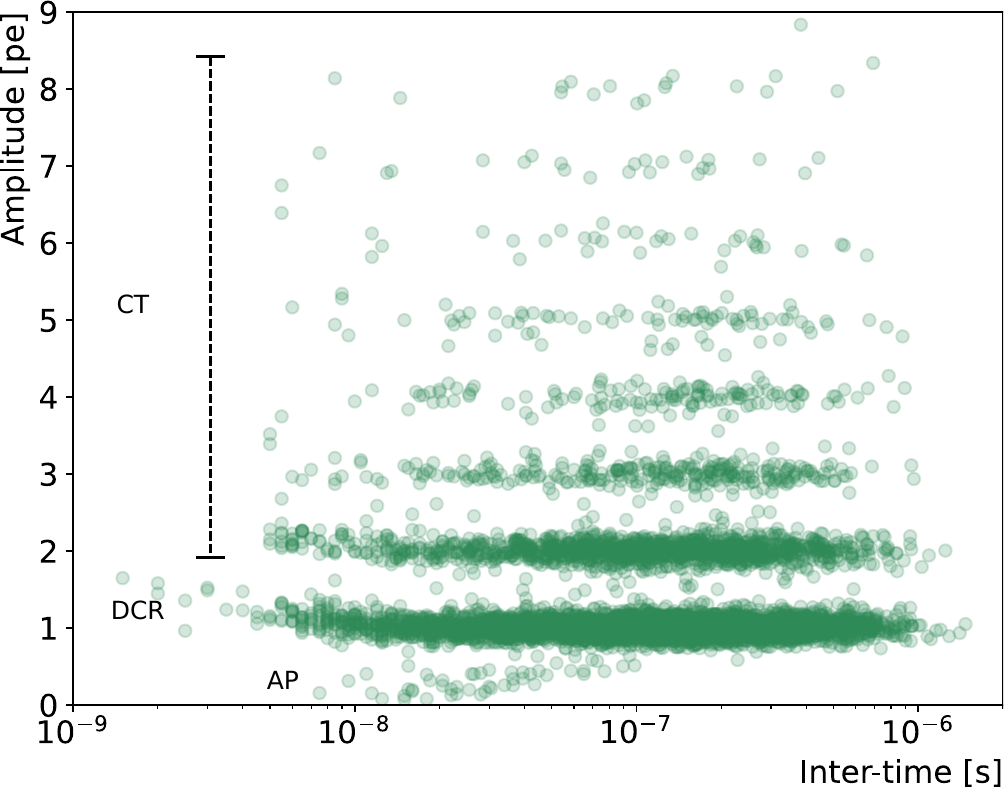}\hspace{0.5cm}
    \includegraphics[width=6.5cm]{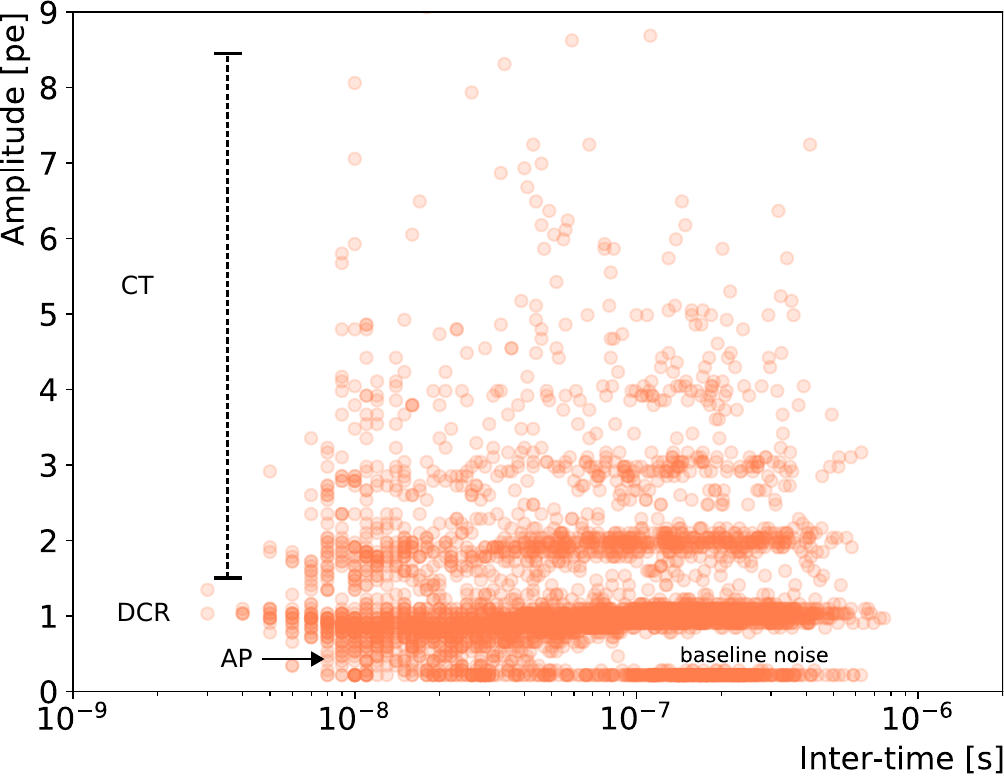}
    \caption{Amplitude vs.\ time difference between consecutive pulses. A comparison of the model (left) and data (right) for the MICROFC-60035 at 32\,V/17$^{\circ}$C. Dark Count Rate (DCR), crosstalk (CT), and afterpulse (AP) components are shown.}
    \label{fig:inter_time}
\end{figure}

The two methods carried out above are amplitude-dependent.  Nevertheless, the model was also evaluated using a time-dependent approach. Figure \ref{fig:inter_time} shows scatter plots of the signal amplitude vs.\ the time difference between consecutive pulses. Both, the model (green-dots) and the data (salmon-dots), indicate an inter-time spread from 3$\times 10^{-9}$\,s to 2$\times 10^{-6}$\,s. The probability of having pulses above 1\,pe decreases with the number of pe and tends to inter-time values between $10^{-7}$\,s to $10^{-6}$\,s. In both cases, contributions of dark counts, crosstalk, and afterpulsing are shown. In the data case, the component $< 0.5$\,pe is contaminated by baseline noise which is not modeled by the simulation algorithm. Additionally, the data quality is affected for values $<$10\,ns due to the sampling resolution of the recording system.

\begin{figure}[ht]
    \centering
    \includegraphics[width=8cm]{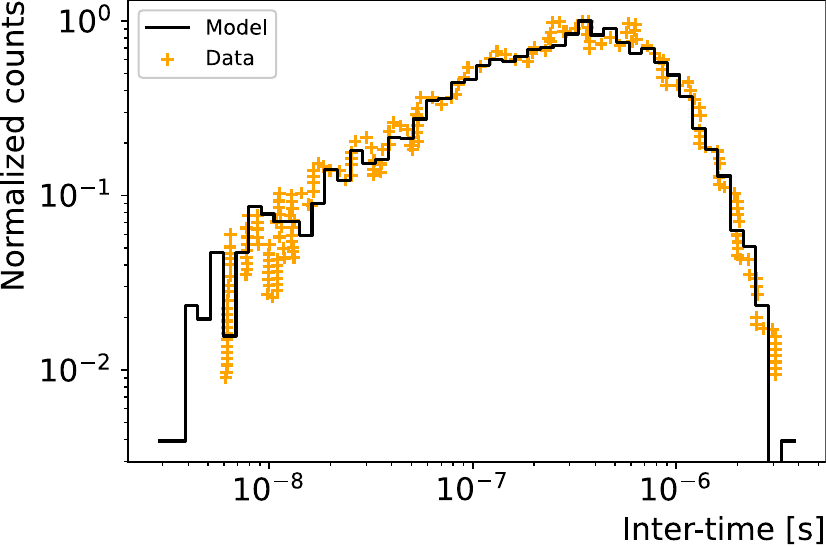}\hspace{0.2cm}
    \caption{Relative time distribution of the NUV-HD3 SiPM at T = 20$^{\circ}$C and V$_{ov}$ = 5.5\,V.}
    \label{fig:psct}
\end{figure}

The model also was validated with the FBK NUV-HD3 SiPM used by the prototype Schwarzschild Couder Telescope (pSCT) at the Cherenkov Telescope Array (CTA) Observatory \cite{Ambrosi2023}. The FBK NUV-HD3 was set to 20$^{\circ}$C and an over-voltage of 5.5\,V. The model input parameters were a DCR of 26\,MHz, a crosstalk of 25$\%$, a recovery of 150\,ns, and a recording window of 100\,$\mu$s. Figure \ref{fig:psct} compares the relative-time distribution between the NUV-HD3 model and the data recorded by the pSCT in dark conditions. The model describes accurately the relative time distribution spanning from $4 \times 10^{-9}$\,ns to $3 \times 10^{-6}$\,ns with the maximum counting at 1/DCR $\sim 3.8 \times 10^{-7}$\,ns as is expected.

\section{Application 1: A Ring Imaging Cherenkov detector}
\label{sec::rich}

\begin{figure}[ht]
    \centering
    \includegraphics[width=14cm]{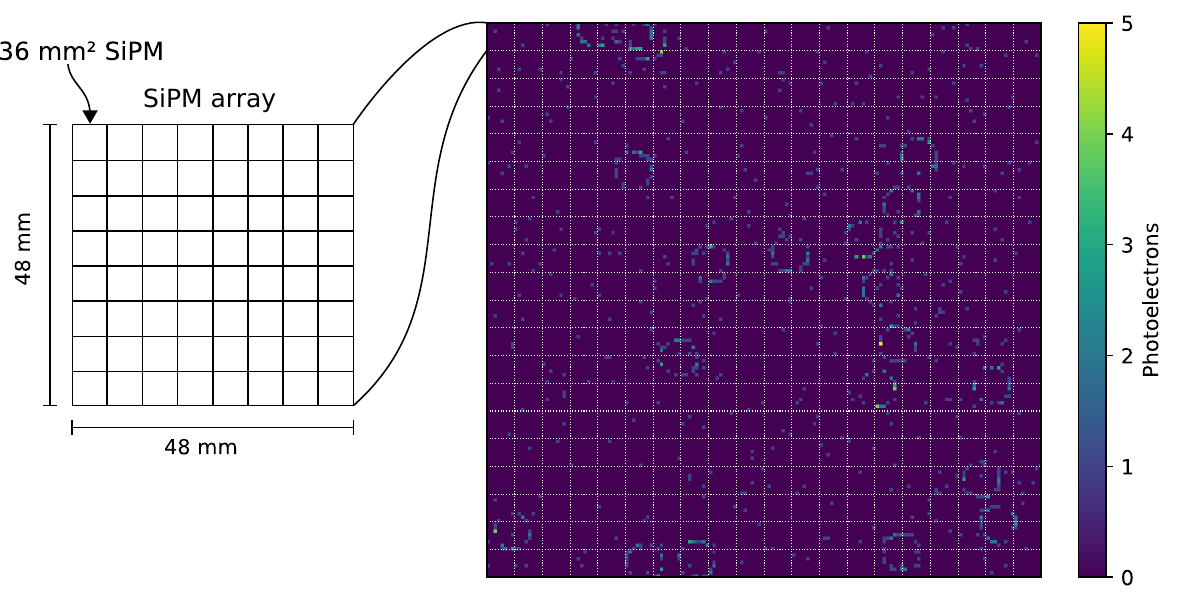}
    \caption{RICH camera made of $20\times 20$ SiPM arrays. The camera detects 20 rings of 27 photons per ring with a 6\,cm radius.}
    \label{fig:camera}
\end{figure}

The camera model was inspired by the RICH detector of the Compressed Baryonic Matter Experiment (CBM) at FAIR, Germany \cite{Klochkov2021}. The CBM RICH camera uses around 1100 H12700 MultiAnode PhotoMultipliers (MAPMTs) arranged in two cameras of 14-rows by 42-columns each. Every H12700 MAPMT comprises 64 pixels (each $6\times6$\,mm$^2$) in an $8\times8$ array \cite{Becker2024}. In this study, we replaced the H12700 MAPMT by a SiPM array made of 64 AFBR-S4N66P024M SiPM. The MaPMT fill factor is 87$\%$ within a total area of 51.5\,mm$\times$51.5\,mm while the fill factor of the SiPM array is 81$\%$ within a total area of 54.1\,mm$\times$52.3\,mm. We simulated the response of the camera with 100$\%$ fill factor as shown in Fig.\,\ref{fig:camera}.

We injected 20 Cherenkov rings of 6\,cm radius and 27 photons/ring on the detection area of the simulated RiCH. The SiPM noise parameters were set to $159\times 10^3$\,Hz/mm$^2$ DCR, 25$\%$ crosstalk, and 1$\%$ after-pulsing (corresponding to being operated at 42\,V/25$^{\circ}$C). The signals generated by the SiPM have a rise time of 2\,ns and a fall time of 3\,ns. The simulation framework was run using a sampling frequency of 2\,GHz. The signal threshold was set to 0.5\,pe for detecting hits in the SiPM array.

\begin{figure}[ht]
    \centering
    \includegraphics[scale=0.45]{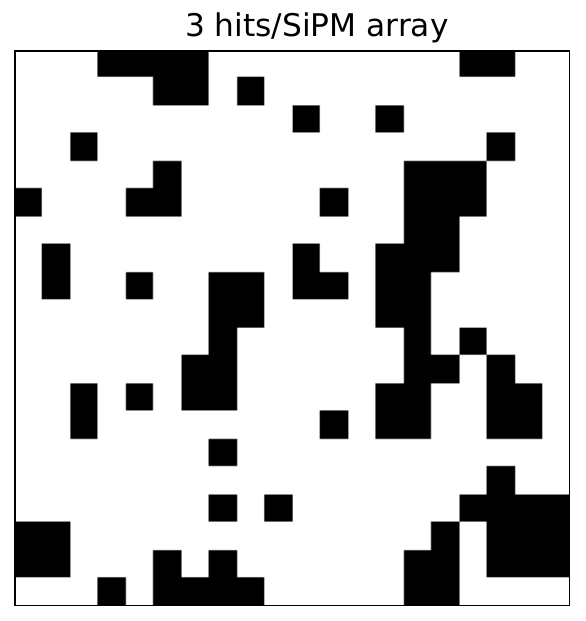}
    \includegraphics[scale=0.45]{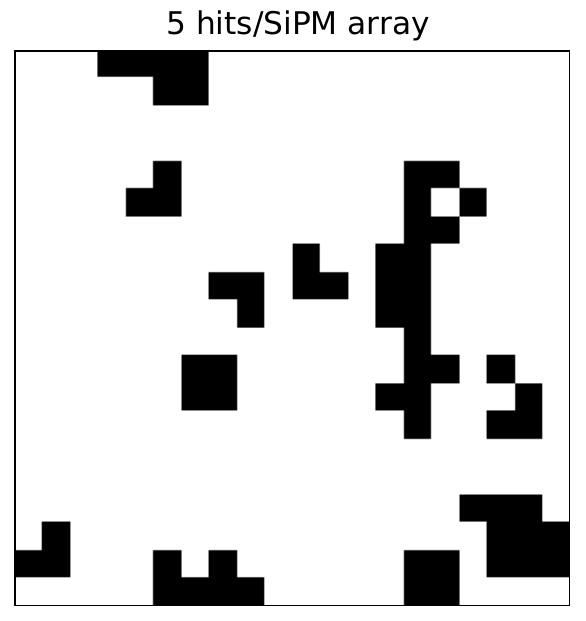}
    \includegraphics[scale=0.45]{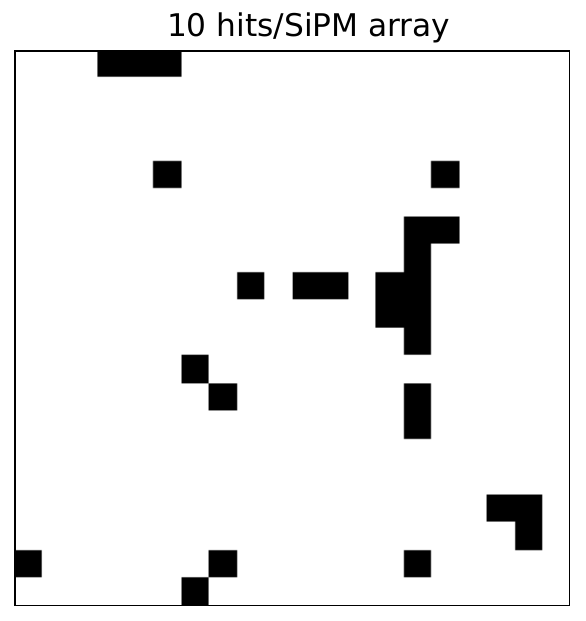}
    \caption{Comparison of the majority voting trigger of the simulated RICH for 3\,hits/SiPM array, 5\,hits/SiPM array, and 10\,hits/SiPM array. Here, each SiPM array is represented by a square. Its colour is black when the corresponding SiPM array is activated, and white when not activated.}
    \label{fig:comparison}
\end{figure}

Figure \ref{fig:camera} depicts the photon camera hits produced by 20 Cherenkov rings and the camera noise within a coincidence window of 5\,ns. The triggering system applies a majority voting threshold activating a SiPM array only if a minimum number of simultaneous hits have been detected on it. Figure \ref{fig:comparison} compares the number of activated SiPM arrays during the event displayed in Fig. \ref{fig:camera} for thresholds of 3\,hits, 5\,hits, and 10\,hits per SiPM array. Each square in the trigger matrix represents an individual SiPM array, when it is fired the square is turned black.

When the threshold is set to 3\,hits per SiPM array, the RICH detects noise as rings, false-positive events. For 5\,hits per SiPM array, all generated rings are well detected, but for 10\,hits per SiPM array some of them are missed. This example shows how this simulation framework can be used to assess the performance of a simple triggering system in a RICH in terms of signal detection efficiency and noise rejection.

\begin{figure}[ht]
    \label{fig:rich_eval}
    \centering
    \includegraphics[scale=0.45]{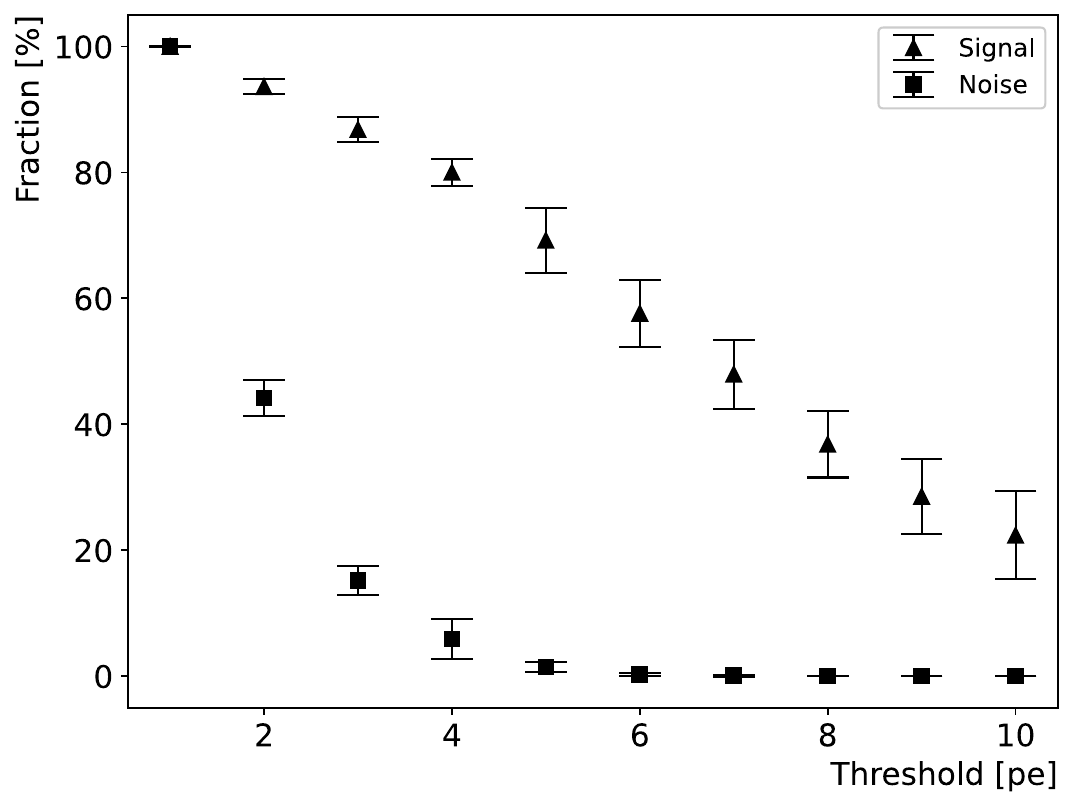}
    \caption{Signal and noise fraction of the RICH as a function of the majority voting threshold.}
    \end{figure}

At the same time, the fraction of rejected noise amounts to 87$\%$ for a threshold of 3 hits per SiPM array, 99.3$\%$ for a threshold of 5 hits per SiPM array, and 100$\%$ hits for the largest threshold of 10 hits per SiPM array. Figure \ref{fig:rich_eval} shows the fraction of signal and noise depending on the applied threshold. The noise is totally rejected above 5 hits per SiPM array preserving 71$\%$ of the signal.

The code used for performing the RICH example is attached in \ref{app::module} and the Python module code is accessible at the \href{https://github.com/JesusPenha/SiPM-APD-MPPC}{Github} repository. Table \ref{tab:parameters} summarizes the model parameters.

\section{Application 2: A Cherenkov telescope}
\label{sec::shower}

\begin{figure}[h]
    \centering
    \includegraphics[width=14cm]{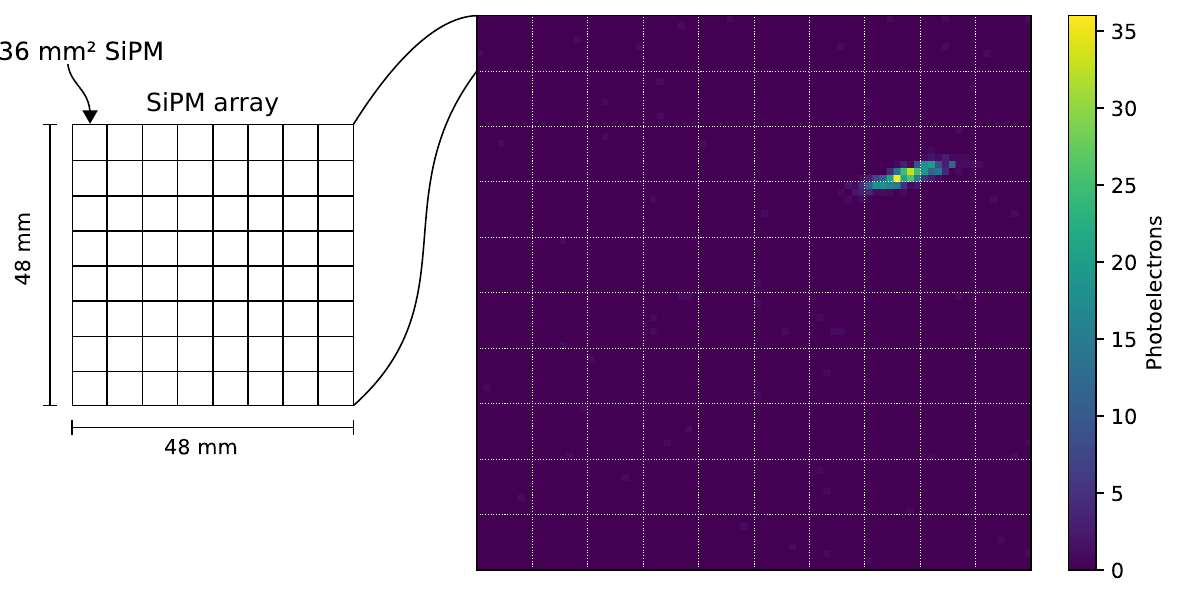}
    \caption{Simulated camera image of an air shower seen by an IACT camera of $10\times 10$ SiPM arrays. The track at the top right corner is caused by 500 photons hitting the camera.  }
    \label{fig:IACT}
\end{figure}

Imaging Air Cherenkov Telescopes (IACTs) are instruments for the detection of atmospheric showers produced by high-energy cosmic rays and gamma rays. Experiments employing IACTs have shown interest in implementing photon detection cameras based on SiPM or APDs. The First G-APD Cherenkov Telescope (FACT) demonstrated the usability of SiPMs in IACTs by implementing a detection camera based on 1440 pixels \cite{Neise2017, Anderhub2011}. One of the Major Atmospheric Gamma-ray Imaging Cherenkov (MAGIC) telescopes has a SiPM-based module installed in one of the imaging cameras to compare its performance with PMT-based modules \cite{Hahn2024}. The Cherenkov Telescope Array Observatory (CTAO) evaluated the performance of a LST (Large Size Telescope) equipped with a SiPM-based camera instead of using PMTs \cite{Berti2020}. Finally, LHAASO (Large High Altitude Air Shower Observatory) plans the realization of ``A Large Array of imaging atmospheric Cherenkov Telescopes" (LACT) based on SiPMs to explore the nature of PeV gamma-ray sources \cite{Lu2024}. A comprehensive recent summary of implementing SiPMs in IACTs can be found in \cite{Ambrosi2022}.  

We simulated an IACT of 10 rows by 10 columns using AFBR-S4N66P024M SiPM operated at 42\,V/25$^{\circ}$C. With this, the telescope camera comprises 100 SiPM arrays each with an array of $8 \times 8$ SiPM, covering a detection area of $48 \times 48$\,cm$^2$.

\begin{figure}[ht]
    \centering
    \includegraphics[scale=0.45]{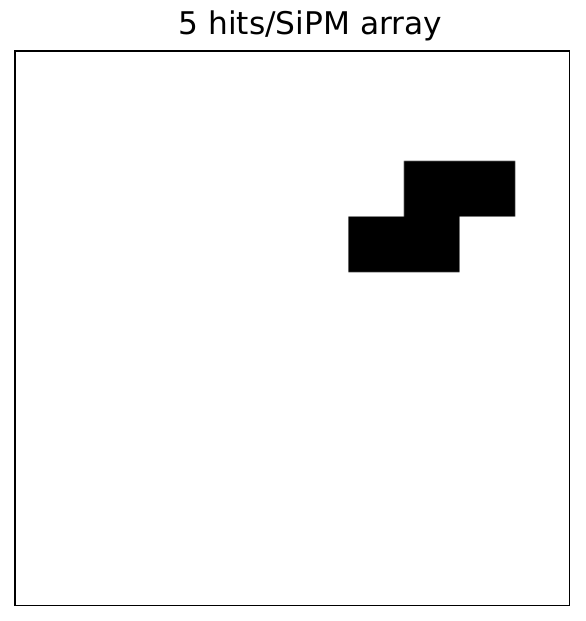}
    \includegraphics[scale=0.45]{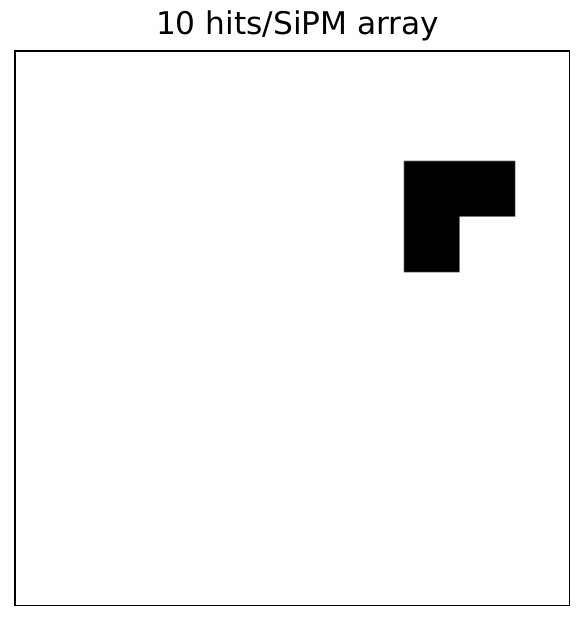}
    \caption{IACT triggering matrix for a discrimination threshold of 5\,hits/SiPM array (left-panel) and 10\,hits/SiPM array (right-panel).}
    \label{fig:shower}
\end{figure}

The simulated shower contains 500 photons impinging the telescope as shown in Fig.\ \ref{fig:IACT}. Figure \ref{fig:shower} compares the telescope triggering matrix applying thresholds of 5\,hits per SiPM array (left-panel) and 10\,hits per SiPM array (right-panel) within a coincidence window of 5\,ns. The shower is completely detected with the 5\,hits per SiPM array threshold, but with the 10\,hits per SiPM array threshold, some shower details are missed.

\section*{Conclusions}

The simulation framework presented in this work performs simulations of Silicon Photomultipliers. The framework models the time behavior of uncorrelated noise (Dark Count Rate) and correlated noise (afterpulsing and crosstalk) of the photosensor from characterization curves depending on temperature and over-voltage. The user can define a large area detector based on SiPM arrays, evaluate the noise contribution, and assess triggering systems based on the expected signal characteristics.

We tested the noise modeling by comparing it with data from four SiPM, AFBR-S4N66P024M, S14160-6050HS, MICROFC-60035, and FBK NUV-HD3. We validated the noise amplitude distribution, the distribution of the relative time between consecutive pulses, and the noise level depending on the number of photoelectrons.

We presented two application examples commonly found in high-energy and astroparticle physics fields. The first was a RICH detector made of $20\times 20$ SiPM arrays, with 64 AFBR-S4N66P024M SiPM each, being hit by 20 Cherenkov rings of 6\,cm radius with 27 photons per ring in a time window of 5\,ns. We evaluated a majority vote trigger system for three thresholds: 3 hits per SiPM array, 5 hits per SiPM array, and 10 hits per SiPM array. The second example was an IACT of $10\times 10$ SiPM arrays made of $8 \times 8$ AFBR-S4N66P024M SiPM each, observing 500 photons originating from an air shower. A majority vote trigger was set at 5 hits per SiPM array and 10 hits per SiPM array. In both the RICH and the IACT, a trigger between 4 to 5 hits/SiPM array provides acceptable performance signal/noise.

\acknowledgments
This work has been supported by ``Netzwerke 2021'', an initiative of the Ministry of Culture and Science of the State of North Rhine-Westphalia.







\bibliographystyle{JHEP}
\bibliography{biblio.bib}

\appendix


\newpage
\section{Using the SiPM Python module}
\label{app::module}

\begin{lstlisting}

# Importing the simulation module

import SiPM_MPPC.sipm as sipm
import matplotlib.pylab as plt
import numpy as np

# Creating a single sipm pulse
# Input parameters

Rt = 2e-9 # Rising time in seconds 
Ft = 50e-9   # Falling time in seconds
A = 1 # Pulse amplitude (pe) photo-electron
R = 0.5 # Time step in ns

pulse = sipm.Pulse(Rt, Ft, A, R, plot=True)
# Output
# pulse, sipm pulse shape with time step R

# Simulating a sipm signal during a recording window
# Input parameters

DCR = 159.6e3  # Dark count rate in Hz/mm2
p_size = 36.0 # SiPM size mm2
CT = 0.31 # Crosstalk normalized to 1
AP = 0.01 # Afterpulse normalized to 1
T_rec = 55e-9 # Recovery time in ns
T_AP = 14.8e-9 # Trapp releasing time in ns
sigma = 0.1 # Amplitude variance in pe
W = 1000  # Recording window in ns
Np = 1 # Number of SiPM

signal, time = sipm.MPPC(pulse, Np, DCR, p_size, CT, AP, T_rec,...
                        ... T_AP, W, R, sigma)
# Output
# signal, sipm signal amplitude in pe
# time, sipm signal time in ns

\end{lstlisting}

\newpage
\begin{lstlisting}

# Generating the peak spectrum and inter-time distributions

A, I, X, Y = sipm.Amplitude_Intertime(signal, N_p, W, R, plot=True)
# Output
# A, amplitude vector in pe
# I, time difference between consecutive pulses in s
# X, peak spectrum x-axis
# Y, peak spectrum y-axis

# Generating the DCR vs. threshold curve
# Input parameters

Lt = 0.1 # Lower threshold in pe
Ut = 8 # Upper threshold in pe
Pt = 200 # Threshold evaluation points

Th, Noise = sipm.DCR_threshold(signal, W, R, Lt, Ut, Pt, plot=True)
# Output
# Th, threshold vector in pe
# Noise, noise frequency in Hz

# Generating noise for a RICH camera made of SiPM arrays
# Input parameters

M = 8 # SiPM array size (M x M)
N_p = M*M # Number of SiPM per array
Nr = 20 # Number of camera rows
Nc = 20 # Number of camera columns
Th = 0.5 # Detection threshold in pe
t0 = 400 # Event trigger time in ns
Cw = 5 # Coincidence window in ns. Detection window (t0 + Cw)

cam_noise = sipm.Camera_noise(signal, Np, p_size, M, DCR, CT, AP, ...
                    ... T_rec, T_AP, W, R, sigma, Nr, Nc, t0, Cw, Th)
# Output
# cam_noise, noise matrix of the RICH camera

\end{lstlisting}

\newpage
\begin{lstlisting}
# Generating photon ring signals

r  = 6.0 # Cherenkov ring diameter in cm
Np_ring = 27 # Number of photons per ring
N_rings = 10 # Number of rings

rings = sipm.Ring_generator(Nr, Nc, r, M, Np_ring, N_rings)
# Output
# rings, signal matrix of the RICH camera

# Plotting camera event

sipm.Camera_plot(rings, cam_noise, M, Nr, Nc)
# Output
# Camera plot

# Evaluating a majority voting trigger

threshold = 3 # Minimum number of photons per SiPM array
sipm.voting_trigger(rings, cam_noise, M, Nr, Nc, threshold,...
                    ...cmap='Greys')

# Output
# Trigger matrix and plot

\end{lstlisting}

\begin{table}[ht]
\caption{Model parameters to be set by the user in simulation set-ups.}
\begin{center}
\begin{tabular}{|l|l|c|c|}
\hline
               \textbf{Model}   & \textbf{Parameters} & \textbf{Symbol} & \textbf{Units}   \\ \hline
\multirow{5}{*}{Pulse} & Rise time & $\tau_R$ &  s  \\  
                  & Fall time & $\tau_F$  &  s  \\ 
                  & Amplitude & $A$ &  pe  \\ 
                  & Time resolution & $R$ &  ns  \\ \hline
\multirow{7}{*}{Noise} & Dark Count Rate & DCR &  Hz/mm$^2$  \\
                  & Crosstalk & CT &  $\%$  \\
                  & Afterpulsing & AP &  $\%$  \\ 
                  & Recovery time & $\tau_{rec}$ &  s  \\ 
                  & Releasing time$^{\ddagger}$  & $\tau_{AP}$ &  s  \\ 
                  & Amplitude variance &  $\sigma_A$ &  pe  \\ 
                  & SiPM Size & $P_{size}$ &  mm$^2$  \\ 
                  & Recording window & $W$ &  ns  \\  \hline
\multirow{7}{*}{Detector} & SiPM array size & $M$ &  -  \\
                  & Number of SiPM per array & $N_p = M \times M$ &  - \\
                  & Number of rows$^{\dagger}$ & $N_r$ &  -  \\ 
                  & Number of columns &  $N_c$  &  -  \\ 
                  & Event trigger time & $t_0$ &  ns  \\ 
                  & Coincidence window* & $C_w$ &  ns  \\ 
                  & Discrimination threshold per SiPM & $Th$ &  pe  \\  \hline
\end{tabular}
\label{tab:parameters}
\end{center}
$^{\ddagger}$ $\tau_{AP}=\tau_{rec}/3.7$ by default\\
$^{\dagger}$Detector size $\rightarrow$ $N_r \times$ $N_c$\\
*Detection window $\rightarrow$ $t_0 + C_w$
\end{table}

\end{document}